\documentclass[aps,amssymb,amsmath,prl,twocolumn,superscriptaddress]{revtex4-1}

\usepackage{graphicx}
\usepackage{amssymb}
\usepackage{epstopdf}
\usepackage{amsmath}
\usepackage[extension=xxx]{hyperref}

\def\be{\begin{equation}}   \def\ee{\end{equation}}
\def\eq#1{{Eq.(\ref{#1})}}    \def\fig#1{{Fig.\ref{#1}}}

\def\kT{k_{{}_{\rm B}}T}     \def\r{{\bf r}}    \def\q{{\bf q}}

\def\Co{{C_o'}}

\newcommand{\corr}[1]{\ensuremath{\langle |#1|^2\rangle}}

\newcommand{\corrt}[1]{\ensuremath{\langle #1(t)  #1(t+\delta t)\rangle}}

\begin{document}

\title{Microphase separation in nonequilibrium biomembranes}
\author{Pierre Sens}
\affiliation{Physico-Chimie Th\'eorique (CNRS UMR 7083), ESPCI, 10 rue Vauquelin, 75231 Paris Cedex 05 - France}
\author{Matthew S. Turner}
\affiliation{Dept. of Physics \& Complexity Centre, University of Warwick, Coventry CV4 7AL, UK.}

\date{\today}



\begin{abstract}

Microphase separation of membrane components is thought to play an important role in many physiological processes, from cell signaling to endocytosis and cellular trafficking. Here, we study how variations in the membrane composition can be driven by fluctuating forces. We show that the membrane steady state is not only controlled by the strength of the forces and how they couple to the membrane, but also by their dynamics: In a simple class of models this is captured by a single a correlation time. We conclude that the coupling of membrane composition to normal mechanical forces, such as might be exerted by polymerizing cytoskeleton filaments, could play an important role in controlling the steady state of a cell membrane that exhibits transient microphase separation on lengthscales in the 10-100 nm regime.
\end{abstract}

\maketitle

Biological membranes are composed of a large number of different lipid and protein species, which are known to phase separate into macrodomains when isolated and cooled below a critical temperature~\cite{keller:1998,veatch:2002,baumgart:2003}. The existence of nano scaled membrane domains at physiological temperature (including so-called {\it lipid rafts}) have been postulated to play an important role in the clustering of signaling proteins~\cite{simons:1997}. The lateral organization of biomembranes has in particular been linked to the structure of the cell cytoskeleton~\cite{kusumi:1996,forgacs:2004,edidin:2006}, and to membrane trafficking. It is now becoming increasingly clear that the controlling principles of biomembrane organisation include the stochastic dynamics of its environment, which influences the membrane through biochemical signaling, mechanical perturbation, and direct material fluxes. Recently we showed how recycling of membrane components can sensitively tune the state of the membrane~\cite{turner:2005}. Here, we analyze the response of the membrane to a fluctuating environment, and show that the level of membrane organization can be tubed by correlations in the dynamical environment. We first calculate the response to a locally correlated stochastic coupling of arbitrary origin, and show how the correlation time of the perturbation controls the membrane's response. We also analyze in detail the case in which membrane composition is coupled to the local membrane curvature, itself driven by random normal forces generated by the cytoskeleton. 

The most physiologically relevant thermodynamic transition in biomembranes is thought to be the liquid disordered (LD) to lipid ordered (LO) transition, which typically separate saturated and unsaturated lipids, and is promoted by cholesterol~\cite{simons:1997}. Since the LO phase exhibits reduced lipid mobility compared to the LD phase, {\em in-vivo} studies of the organization of biological  membranes can emploit an approach based on dynamical tracking of tagged membrane component~\cite{kusumi:1996}. A particularly interesting example~\cite{serge:2008} demonstrates the existence of short-lived ($\sim250$msec.) low mobility nano-domains ($\sim75$nm), appearing and disappearing near seemingly well-defined positions on the membrane. Such behaviour could result from biochemical or physically localized stochastic forces that act to drive changes in membrane composition. 

\noindent{\bf Model for membrane phase separation}. Our starting point is to write down a Hamiltonian in terms of the membrane composition field $\rho(\r)$ and normal membrane displacement $u(\r)$, both assumed to represent small perturbations around a perfectly mixed ($\rho=0$) and flat ($u=0$) membrane~\cite{leibler:1986}:
\begin{multline}
{\cal H}=\frac{1}{2}\int d^2\r\left(b\rho^2+\mu(\nabla\rho)^2\right.\\ 
\left.+\kappa(\nabla^2 u)^2+\sigma(\nabla u)^2-2\kappa \Co\rho\nabla^2 u-2\zeta\rho-2fu\right)
\label{H}
\end{multline}
which can either be motivated on symmetry grounds or, by tracing the physical interpretation of the various terms. These include: a local mixing interaction and concentration gradient term for the membrane composition (parameters $b$ and $\mu$), the membrane tension $\sigma$ and bending rigidity $\kappa$, and a coupling term $\Co$ capturing how the local membrane spontaneous curvature depends on the composition to lowest (linear) order in $\rho$. Thus $\rho$ is the density of membrane component(s) that couple to curvature. The fields $f$ and $\zeta$ are used to capture the coupling of external (cellular) forces conjugate to $u$ and $\rho$ respectively. In the following, we will mostly work in the spatial Fourier space, with $x_\q=\int d^2\r e^{i\q\cdot\r}x(\r)$, giving ${\cal H}=\int\frac{d^2\q}{2(2\pi)^2}{\cal H}_q$ with
\begin{eqnarray}
&{\cal H}_q=h_q|\rho_q|^2+k_q|u_q|^2-2\beta_q u_q\rho_{-q}-2\zeta_q\rho_{-q}-2f_q u_{-q}\cr
&h_q=b+\mu q^2\qquad k_q=\sigma q^2+\kappa q^4\qquad \beta_q=\kappa \Co q^2
\end{eqnarray}

\begin{figure}[b] 
   \includegraphics[width=8.5cm]{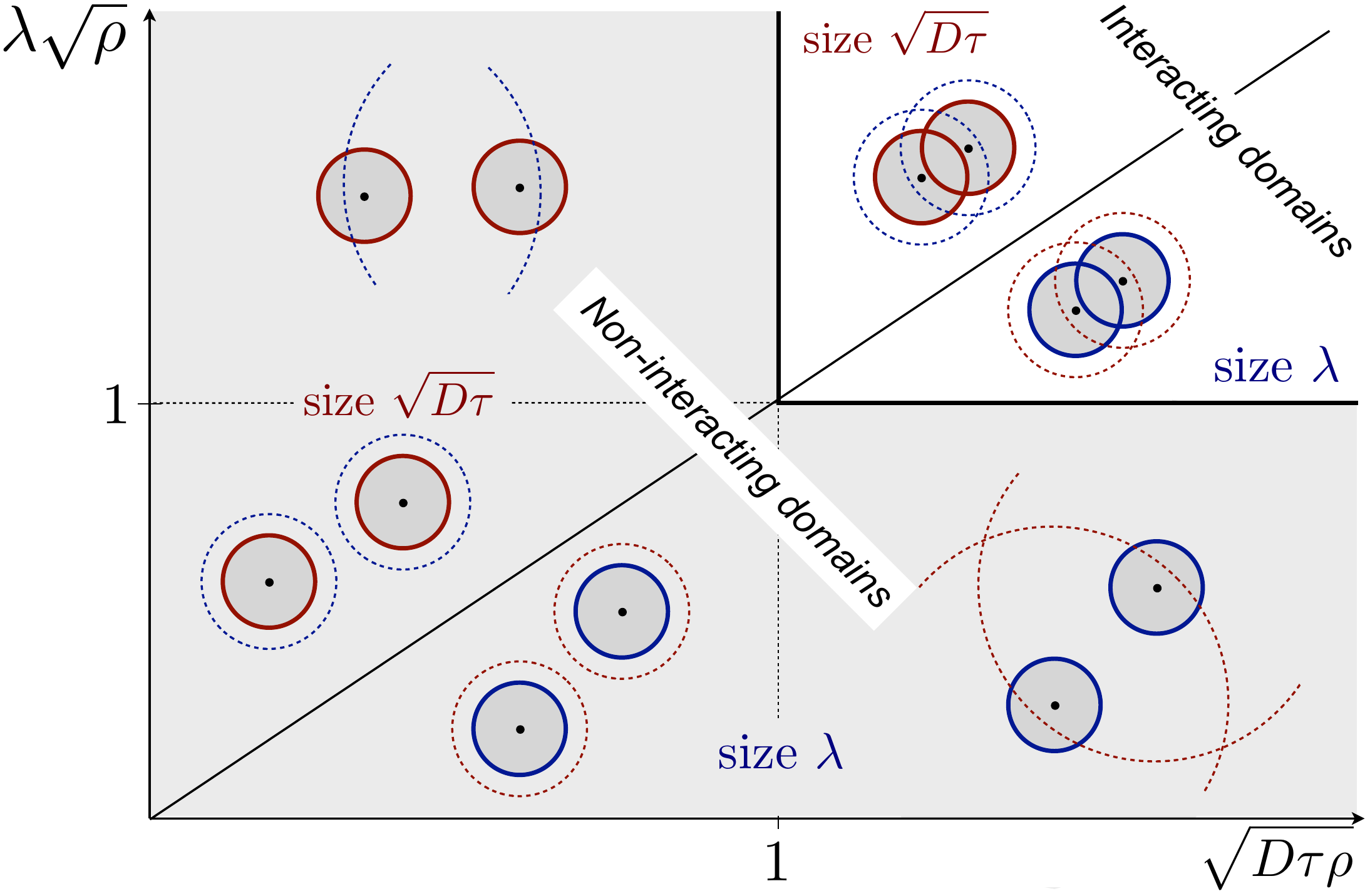} 
   \caption{The dynamics of a membrane containing multiple force centres (of density $\rho$) depends on the ratio  of the membrane length $\lambda$ (see text) to the distance between force centres:  $\lambda\sqrt\rho$, and the ratio of the composition diffusion length $\sqrt{D\tau}$ to the distance between force centres  $\sqrt{D\tau\rho}$. As we discuss in the text independent force centres each contribute linearly to the membrane displacement and enrichment. Each cartoon represents two neighbouring force centres, with a typical separation $\rho^{-1/2}$, and shows the diffusion length in red, the membrane length in blue (not to scale) and the region of substantial compositional variation in dark grey shading. In each case the controlling lengthscale is shown as a continuous coloured line and the other lengthscale is shown dotted. The shaded regimes shows where domains are substantially non-interacting.}
\label{phasefig}
\end{figure}

The kinetic evolution of the two fields, subjected to local fluctuating forces $f(t)$ and $\zeta(t)$ coupling to the displacement and composition, respectively, is given by the equations~\cite{chaikin:1995}:
\begin{eqnarray}
\dot\rho_q+\Lambda q^2(h_q\rho_q-\beta_qu_q)&=&\Lambda q^2\zeta_q\cr
\eta q \dot u_q+(k_q u_q-\beta_q\rho_q)&=&f_q
\label{kineqrhou}
\end{eqnarray}
where $\Lambda$ is related to the diffusion coefficient of membrane component by $D=\Lambda h_q$ ($D_o=D(q\to 0)$), and where $\eta$ is the solvent viscosity dampening the membrane displacement~\cite{reister:2010}. Our goal is to study the effect of correlated fluctuating forces (of zero mean) on the membrane state, to which end we adopt
\begin{eqnarray}
\langle\zeta_q(t)\rangle=0&\quad&\langle\zeta_q(t)\zeta_{-q}(t+\delta t)\rangle=\bar\zeta_q^2e^{-|\delta t|/\tau_\rho}\cr
\langle f_q(t)\rangle=0&\quad&\langle f_q(t) f_{-q}(t+\delta t)\rangle=\bar f_q^2 e^{-|\delta t|/\tau}
\label{force}
\end{eqnarray}
In the case of more than one localised force we can write $f(\r,t)=\sum_i f_i(\r-\r_i,t)$, with $f_i$ the force due to the $i^{\rm th}$ cytoskeletal anchor; similarly for $\zeta$ where its canonical force would be due to some localised biochemical coupling(s). Thus, with $f_i(\r-\r_i,t)=f_i(t)\delta(\r-\r_i)$ (here we consider applied forces $f$ with zero range spatial correlations) $f_q=\sum_i e^{iq\cdot\r_i} f_i(t)$ with $\langle f_i(t)\rangle=0$ and $\langle f_i(t)f_j(t+\delta t)\rangle=\bar f^2\delta_{ij} e^{-\delta t/\tau}$. Here $\bar f$ is simply a constant. In what follows we restrict our attention to {\it single force centres} situated at $\r=0$ noting that the membrane response is {\it linear} and so membrane deformation caused by additional force centre(s) would simply add independently. For a numerical solution of the microphase separation induced by multiple force centres see supporting movies~1-4 while the level of interactions between these force centres is shown in \fig{phasefig}.

\noindent{\bf Fluctuations of membrane composition.} We first disregard the coupling between membrane composition and deformation ($\beta=0$) and study the fluctuations of membrane composition subjected to a correlated fluctuating force $\zeta(t)$. The solution of \eq{kineqrhou} subjected to \eq{force} can then be calculated (see Supplementary Informations - S.I.):
\be
\corr{\rho_q}=\frac{\bar\zeta^2}{h_q^2}\frac{Dq^2\tau_\rho}{Dq^2\tau_\rho+1}
\label{corrrho}
\ee
This expression gives the appropriate static response ($\rho_q=\zeta_q/h_q$) in the limit $Dq^2\tau\gg1$, and satisfies equipartition of energy: $\corr{\rho_q}=\kT/h_q$ in the limit of temporally uncorrelated forces ($\tau_\rho\rightarrow0$) if the fluctuating force $\zeta(t)$ satisfies the fluctuation-dissipation theorem: $\corr{\zeta_q}=\bar\zeta_q^2=\kT/(\Lambda q^2\tau_\rho)$. \eq{corrrho} shows that modes satisfying $Dq^2\tau_\rho\gg1$ are essentially unaffected, while modes such that $Dq^2\tau_\rho\ll1$ are strongly suppressed,  by the fluctuating nature of the correlated force.

\begin{figure*}[bt] 
   \centering
   \includegraphics[width=18cm]{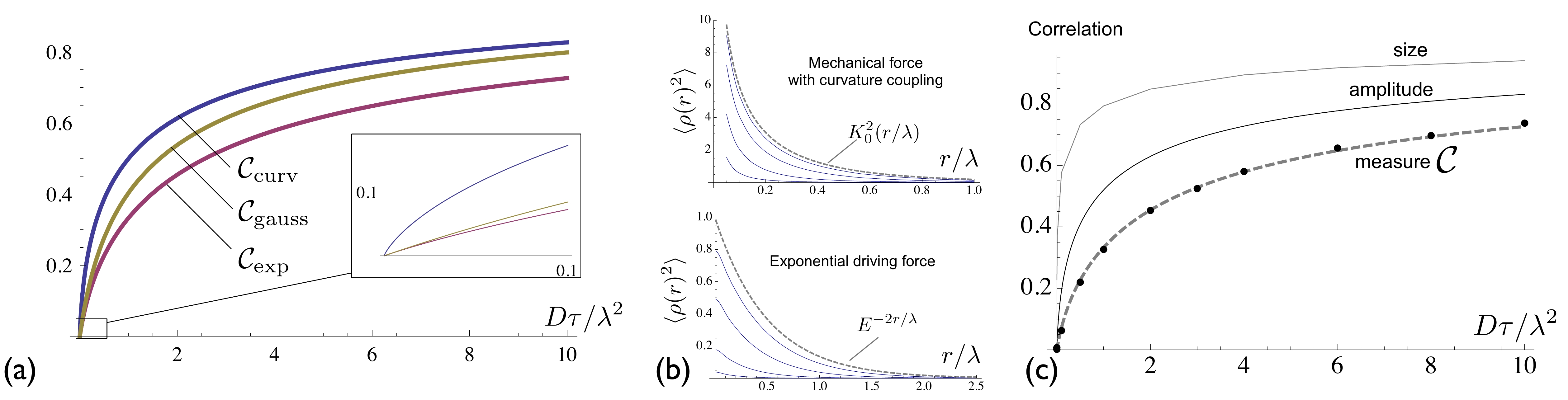} 
   \caption{{\bf a} Comparizon of the membrane dynamical response (as defined by \eq{s}) to a fluctuating environment of typical correlation time $\tau$ that drives the membrane composition changes via {(i)} localized exponentially decreasing forces ($s_{\rm exp}$), {(ii)} localized forces with guassian spatial distribution ($s_{\rm gauss}$) and {(iii)} a local coupling between membrane composition and membrane curvature, itself excited by localized normal mechanical forces ($s_{\rm curv}$).  The inset shows the different behaviors of the response functions for weak correlations: $\tau\rightarrow0$. {\bf b.} Local correlation function $\langle\rho(r)^2\rangle$ driven by a fluctuating random force with a correlation time $\tau$ (shown for $\tau=0.01,\ 0.1,\ 1,\ 5,\ 10$). Left shows the case of a local mechanical driving force with coupling between membrane composition and curvature, Right: shows the case of a force decaying exponentially from a local source. In dashed is shown the expected result for an infinitely correlated (permanent) force. {\bf c.} Evolution of the amplitude at the origin $\langle\rho(0)^2\rangle$ (black) and the distance $r_{1/2}$ from the origin where the amplitude has dropped by half (grey) for an exponential perturbation, as a function of the perturbation correlation time $\tau$. The global measure ${\cal C}$ (dashed grey) is numerically shown to follow the expected scaling behaviour ${\cal C}=\langle\rho(0)^2\rangle r_{1/2}^2$ (dots)}
\label{results}
\end{figure*}

\noindent{\bf Membrane-cytoskeleton coupling}. The cytoskeleton underlying the plasma membrane has long been thought to influence the mobility and clustering of membrane proteins. So far, attention has mostly been directed toward the role of cytoskeleton filament lying parallel to the membrane, thereby creating hindrance to the lateral motion of membrane proteins (the so-called {\it picket-fence} model)~\cite{kusumi:2005}.  The membrane composition is however also known to be sensitive to the local membrane conformation (mostly its curvature~\cite{parthasarathyab:2007}), which should fluctuate strongly under the action of mechanical forces produced by polymerizing and depolymerizing cytoskeleton filaments. 
This situation can be mathematically addressed through the parameter $\beta$ coupling the local membrane composition to its local curvature.  One can in principle solve \eq{kineqrhou} for the two coupled fields of membrane composition and deformation  (see for instance~\cite{chaikin:1995}). For the sake of clarity here we only investigate the limit of rather weak coupling ($\beta_q^2<k_qh_q$), so that the membrane deformation is substantially determined by the external force, with little feedback from the membrane composition. It is known that, in the opposite limit of strong coupling, a curvature instability can arise where large membrane undulations coupled to strong composition variation grow spontaneously~\cite{leibler:1986}. 

The Fourier transform of the curvature of a membrane deformed by a static point force is $ q^2u_q=f/(\sigma(1+(\lambda q)^2))$, with $\lambda\equiv\sqrt{\kappa/\sigma}$ the characteristic ``healing length'' of membrane deformation ($\sim 100nm$ for $\kappa=20\kT$ and $\sigma=10^{-5}J/m^2$). In real space, the curvature is proportional to the modified Bessel function  of the second kind $K_0(r/\lambda)$~\cite{evans:2003}, decaying exponentially at large distance, but diverging logarithmically at short distance. Solutions must be regularized by introducing a short-length cut-off $a$, effectively corresponding to the (molecular) lateral size over which the point force acts (we assume $a\ll\lambda$). This linear description is valid in the small deformation limit $\nabla u\ll1$, requiring that the strength of the point force is smaller than $\sqrt{\kappa\sigma}\sim 10-20$pN, beyond which the membrane deforms into a tubular spicule. 

The mean squared curvature solution of \eq{kineqrhou} under the fluctuating force $f$ satisfying \eq{force} is (see S.I.):
\be
\corrt{u_q}=\frac{\bar f^2}{k_q^2}\frac{(\alpha_q \tau)^2}{(\alpha_q \tau)^2-1}\left(e^{-\delta t/\tau}-\frac{e^{-\alpha_q\delta t}}{\alpha_q\tau }\right)\
\label{correlu}
\ee
where $\alpha_q=k_q/(\eta q)$ is the relaxation rate of the membrane deformation~\footnote{Thermal equilibrium requires  $\tau\rightarrow0$ and   $\bar f^2=2\kT\eta q$}. As a result of its coupling to the local curvature, the membrane composition sees an additional fluctuating force with  correlation $\corrt{\zeta_q'}=\beta_q^2\corrt{u_q}$ coming from the  fluctuations of the membrane shape.

The composition correlation function now involves three relaxation rates: of the force correlation $1/\tau$, of the membrane deformation $\alpha_q$, and of the membrane composition $Dq^2$. In the regime of most interest to us the membrane shape adjusts much faster than the membrane composition $\alpha_q\gg Dq^2$, which is the case provided the membrane length $\lambda$  is much smaller than $\kappa/(D\eta)(\sim 10\mu$m). This is always the case in practice ($\lambda\sim 10^{-2}-1\mu$m depending on membrane tension), the membrane deformation can therefore be considered to relax instantaneously to its mechanical equilibrium on the timescale of composition changes. The correlation functions then reads:
\be
\corr{\rho_q}=\frac{(\lambda \Co)^2\bar f^2}{h_q^2(1+(\lambda q)^2)^2}\frac{Dq^2\tau}{(Dq^2\tau+1)}
\label{corrrho2}
\ee
akin to \eq{corrrho} but with an effective driving force that includes the membrane elasticity and the strength $C_0'$ of the composition coupling to curvature but which is insensitive to any details regarding the dynamics of its deformation.

\noindent{\bf Quantification of membrane organization}. The fulll spatio-temporal correlation function for the membrane composition (see S.I.) can, in principle, yield any statistical information of the membrane composition, e.g. by inverse transformation to real space (\fig{results}b). This step typically involves numerical calculation so it is desireable to also obtain an analytic statistical measure for the membrane state. Towards this end we define the integrated autocorrelation function:
\be
{\cal C}(\tau_\rho)=\frac{\int d^2\r\corr{\rho(\r)^2}}{\int d^2\r \bar\zeta^2/h_q^2}=\frac{\int d^2q \frac{\bar\zeta^2}{h_q^2}\frac{D\tau_\rho q^2}{D\tau_\rho q^2+1}}{\int d^2q\bar\zeta^2/h_q^2}
\label{s}
\ee
here normalized by the membrane's response to a permanent perturbation (${\cal C}(\tau_\rho\rightarrow\infty)=1$) to reveal the role of the fluctuating nature of the force. In principle, the susceptibility of the membrane composition $h_q=b+\mu q^2$ also contains the static composition correlation length $\sqrt{\mu/b}$. We consider the membrane to be far from the critical point for membrane composition, in which case this length scale is of molecular dimension~\cite{chaikin:1995} and we may use the approximation $\mu=0$ henceforth.

\eq{s} can readily be used to quantify the membrane composition's response to fluctuations of the local curvature under stochastic mechanical forces (using \eq{corrrho2}):
\be
{\cal C}_{\rm curv}=\frac{\bar\tau(\bar\tau-\log\bar\tau-1)}{(\bar\tau-1)^2}
\label{scurv}
\ee
with $\bar\tau\equiv D\tau/\lambda^2$. This results can be compared to the response of other types of fluctuating environments, such as an exponentially decreasing local source of signaling molecules (with $\bar \zeta_q=1/(1+(\lambda q)^2)^{3/2}$):
\be
{\cal C}_{\rm exp}=\frac{\bar\tau(\bar\tau^2-2\bar\tau\log\bar\tau-1)}{(\bar\tau-1)^3}
\label{sexp}\ee
or a similar force of Gaussian form: $\bar \zeta_q=e^{-(\lambda q)^2/2}$. The corresponding response functions are shown in \fig{results}a. All three show a  slow convergence toward the static response (${\cal C}\rightarrow 1$  for $\bar\tau\rightarrow\infty$), but exhibit quite distinct behaviour for weakly correlated excitations (small $\bar\tau$):
\be
{\cal C}_{\rm curv}\xrightarrow[\bar\tau\rightarrow0]{}\bar\tau\log\frac{1}{\bar\tau}\qquad {\cal C}_{\rm exp}\sim {\cal C}_{\rm gauss}\xrightarrow[\bar\tau\rightarrow0]{}\bar\tau
\label{scompare}
\ee
This difference comes from the fact that the membrane's conformational response to a localized mechanical force contains a high density of short wavelength modes to which the membrane composition is quick to adjust by diffusion, thereby leading to a logarithmically diverging sensitivity to correlations of the stochastic environment for ${\cal C}_{curv}$. Note that  Eqs.(\ref{scurv},\ref{sexp},\ref{scompare}) are valid provided $\tau> a^2/D$, which is a weak constraint since the short-distance cutoff $a$ is of molecular size.

Two important measures of the fluctuating membrane ``domains'' are the intensity of phase segregation within them, and their average size. Both properties depends on the correlation time $\tau$, and are somehow mixed in the global measure ${\cal C}$  of \eq{s}. As show in \fig{results}b, the real-space correlation function $\langle\rho(r,t)^2\rangle$ can be obtained from the (numerical) inverse Fourier transform of $\langle \rho_q(t)\rho_{q'}(t)\rangle$ (see S.I., Eqs.(5-6)). These plots show that while the characteristic shape of the domain resembles the permanent shape even for short correlation time $\tau$ (of order $\sim \lambda^2/D$), the amplitude of the reorganization requires longer times to reach its static level. This behaviour is further illustrated in \fig{results}c, where both the maximal level of correlation ($\corr{\rho(a)}$) and the typical distance $r_{1/2}$  from the source at which the amplitude falls by half ($\corr{\rho(r_{1/2})}=\corr{\rho(a)}/2$), are plotted as a function of the correlation time. Remarkably, the global measure of the perturbation ${\cal C}$  closely  follows the expected scaling result ${\cal C}\sim\langle\rho(a)^2\rangle r_{1/2}^2$ for all times, thereby validating ${\cal C}$  as a good quantity with which to measure the membrane's response to a fluctuating environment\footnote{\fig{results}c presents the case of an exponential perturbation only, has such measures depend on the sort-distance cut-off $a$ in the case of a perturbation by mechanical force, since the response diverges exponentially at short distance. This complication does not modify the validity of \eq{scurv} provided $\tau>a^2/D$}.

\noindent{\bf Conclusion}. The processes that control the lateral organization of biological membranes are still far from being well understood. Much effort has been put into understanding the equilibrium phase behaviour of these membranes as part of a philosophy that sees them as quasi-equilibrium structures. We believe that active fluctuations in the membrane's environment play a crucial role in this microphase separation and that this facilitates the sensitive transmembrane signalling of environmental variations. Our model shows the crucial importance of the temporal correlation of a noisy environment in triggering the membrane response. \fig{phasefig}

 \fig{results} shows the relative strength of the response with the correlation time of the perturbation (compared an infinite correlation time). While the extent of the membrane reorganization reaches the extent of the driving force even in a highly fluctuating environment (for correlation times larger than $\lambda^2/D$, of order $100$ msec for $\lambda=100$nm and $D=0.1\mu $m$^2$/s), the amplitude of the perturbation must be driven much more slowly if it is to reach its full potential. A primary result of this work is that compositional changes due to variations in membrane curvature driven by normal forces exerted by the cytoskeleton could represent the dominant cellular strategy for membrane organization. 

\begin{acknowledgments}
We thank the french ANR (P.S.)  for financial support, and we thank the KITP (UCSB, Santa Barbara) where this work was initiated.
\end{acknowledgments}

\end{document}


\title{Supporting Information for {\em ``Microphase separation in nonequilibrium biomembranes''}}
\author{Pierre Sens \& Matthew S. Turner}

\centerline{{\Large\bf Supporting Information}}
\vspace{0.5cm}
\centerline{\large ``Microphase separation in nonequilibrium biomembranes''}
\vspace{0.5cm}
\centerline{Pierre Sens \& Matthew S. Turner}

\vspace{5mm}

Below, we detail the calculation of the correlation functions for the membrane composition and deformation appearing in the main text (Eqs.(5,6)). We first define the integral:
\be
{\cal I}_{\alpha,\alpha'}(t,t+\delta t)=\int_0^t \alpha dt_1\int_0^{t+\delta t}\alpha'dt_2 e^{-|t_1-t_2|/\tau}e^{-\alpha(t-t_1)}e^{-\alpha'(t+\delta t-t_2)}
\label{I}
\ee
which has the asymptotic behaviour
\begin{eqnarray}
{\cal I}_{\alpha,\alpha'}&\xrightarrow[t\rightarrow\infty]{} &\frac{\alpha'\tau}{\alpha'\tau-1}\left(\frac{\alpha\tau}{\alpha\tau+1}e^{-\delta t/\tau}-\frac{2}{(\alpha'\tau+1)(1+\alpha'/\alpha)}e^{-\alpha'\delta t}\right)\cr
{\cal I}_{\alpha,\alpha}&\xrightarrow[t\rightarrow\infty]{}&\frac{\alpha\tau}{((\alpha\tau)^2-1)}\left(\alpha\tau e^{-\delta t/\tau}-e^{-\alpha\delta t}\right)
\label{I2}
\end{eqnarray}

We are interested by the correlation function of a variable satisfying a dynamical equation of the form (Eq.3, main text)
\begin{eqnarray}
\gamma_q\dot\phi_q+k_q\phi_q=\xi_q\quad {\rm with}\quad  \langle\xi_q\rangle=0\ ;\ \langle\xi_q(t)\xi_{q'}(t+\delta t)\rangle=\bar\xi_{qq'}e^{-|\delta t|/\tau}
\label{kineq}
\end{eqnarray}

The time dependent solution of this equation reads:
\be
 \phi_q(t)=\phi_q(0)e^{-\alpha_q t}+\int_0^t\alpha_qdt'\frac{\xi_q(t')}{k_q}e^{-\alpha_q|t-t'|}\qquad\alpha_q\equiv\frac{k_q}{\gamma_q}
 \ee
yielding the correlation function 
\begin{eqnarray}
\langle \phi_{q}(t)\phi_{q'}(t+\delta t)\rangle&\xrightarrow[t\rightarrow\infty]{}&\frac{\bar\xi_{qq'}}{k_qk_{q'}}{\cal I}_{\alpha_q,\alpha_{q'}}(t,t+\delta t)
\label{A-correl}
\end{eqnarray}
from which the results given in Eqs(5,6) can easily be derived.

The membrane response to a point force being radially symmetric, the real-space correlation functions (such as those shown in Fig.2 of the text) can be calculated with
\be
\langle \phi(r,t)\phi(r', t+\delta t)\rangle=\int 2\pi q dq J_0(qr)\int 2\pi q' dq' J_0(q'r')\langle\phi_q(t)\phi_{q'}(t+\delta t)\rangle
\ee
where $J_0$ is the Bessel function of the first king, satisfying $\int_0^{2\pi}d\theta e^{iq.r\cos\theta}=2\pi J_0(q.r)$

\begin{figure}[t] 
 \includegraphics[width=15cm]{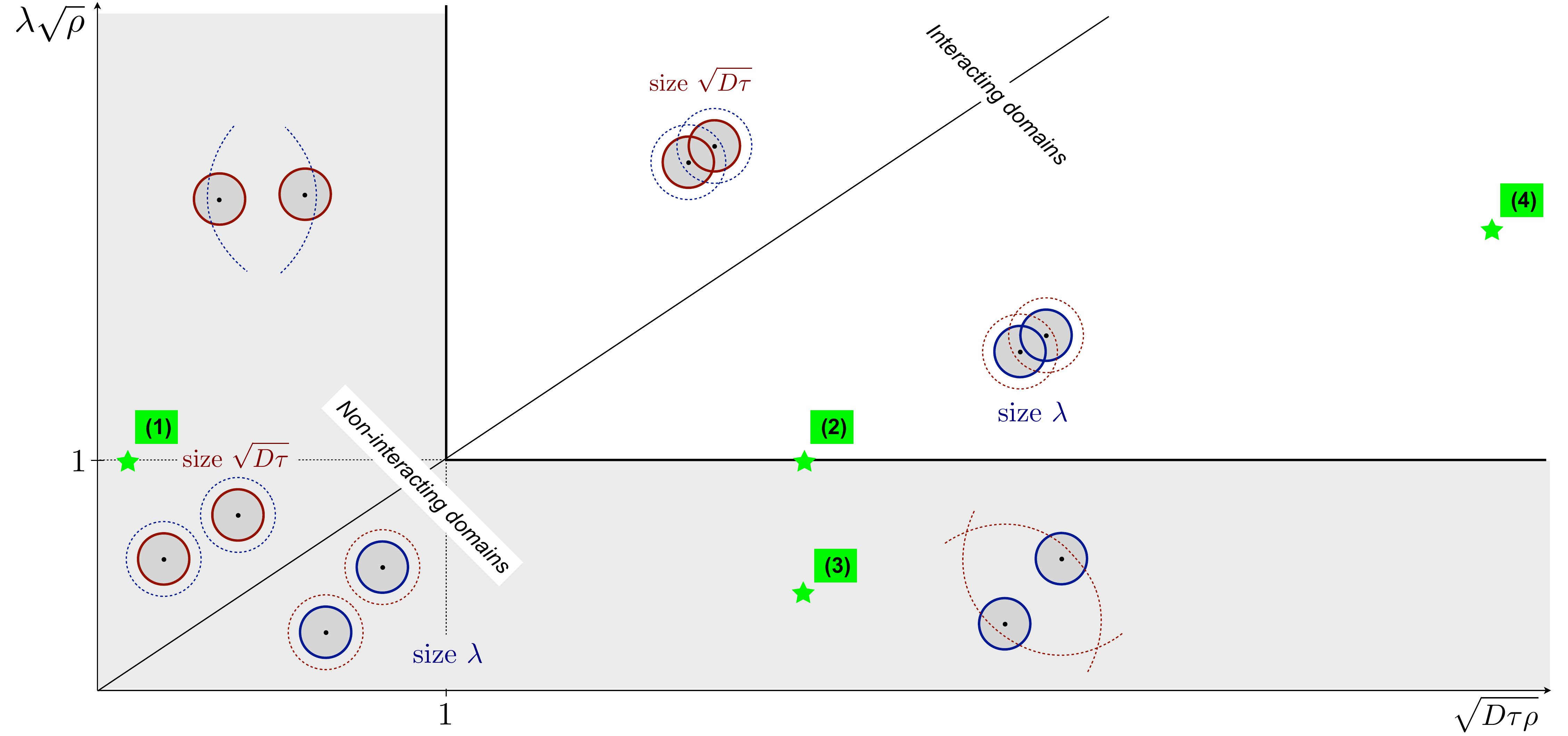} 
   \caption{Reproduction of the phase diagramm appearing in the text (Fig.1) with indication of the states represented in the different supplementary movies (green stars, number corresponds to the movie numbers).}
   \label{fig:example}
\end{figure}